\documentclass[aps,pre,reprint]{revtex4-1}

\usepackage[pdftex]{graphicx}
\usepackage{amsmath}
\usepackage{amsfonts}
\usepackage{mathrsfs}
\usepackage{bm}
\usepackage{hyperref}

\begin{document}

\title{Smallest neural network to learn the Ising criticality}

\author{Dongkyu Kim}
\affiliation{Department of Physics and Photon Science, School of Physics and Chemistry, Gwangju Institute of Science and Technology, Gwangju 61005, Korea}
\author{Dong-Hee Kim}
\email{dongheekim@gist.ac.kr}
\affiliation{Department of Physics and Photon Science, School of Physics and Chemistry, Gwangju Institute of Science and Technology, Gwangju 61005, Korea}

\begin{abstract}
Learning with an artificial neural network encodes the system behavior in a feed-forward function with a number of parameters optimized by data-driven training. An open question is whether one can minimize the network complexity without loss of performance to reveal how and why it works. Here we investigate the learning of the phase transition in the Ising model and find that having two hidden neurons can be enough for an accurate prediction of critical temperature. We show that the networks learn the scaling dimension of the order parameter while being trained as a phase classifier, demonstrating how the machine learning exploits the Ising universality to work for different lattices of the same criticality within a single set of trainings in one lattice geometry.
\end{abstract}

\maketitle

\section{Introduction}

Machine learning~\cite{Hinton2006,LeCun2015} is a framework for prediction based on data-driven optimization of a hidden complex structure of unknowns, which drastically differs from a conventional model of explanation based on a physical understanding of a system. Performing as a classifier, an artificial neural network can suggest a proper label for an unacquainted input without doing explicit analysis, which is done by training a large set of the network parameters to adapt themselves to already labeled data. In spite of many empirical successes, one intrinsic issue is that the network often works like a ``black box'' since it is generally difficult to see inside how it reaches a particular output. Such lack of transparency is due to the high complexity coming out of the interplay between many network parameters. The more complex structure may help increase flexibility in learning but at the same time makes it harder to understand how it extracts a desired feature from the data. In this paper, we present the opposite extreme of a minimally simple neural network to explain the observed accuracy and universality in its learning of the phase transition in the Ising model.

The ideas of machine learning have been actively applied to problems in classical and quantum physics. For instance, efficient Monte Carlo simulation methods were proposed by integrating machine learning into wave-function representations~\cite{Carleo2017,Nomura2017,Gao2017,Deng2017,Cai2018,Glasser2018,Chen2018,Torlai} and cluster updates~\cite{Huang2017,Wang2017,Liu2017a,Liu2017b,Xu2017,Nagai2017}. On the other hand, phase transitions have been extensively examined in various schemes of the supervised~\cite{Carrasquilla2017,vanNieuwenburg2017,Broecker2017,Chng2017,Schindler2017,Tanaka2017,Wetzel2017b,Zhang2017a,Zhang2017b,Zhang2018,Beach2018,Suchsland2018,KochJanusz2018,Iakovlev,VargasHernandez,Hsu,Dong} and
unsupervised~\cite{Wang2016,Torlai2016,Hu2017,Costa2017,Wetzel2017a,Ponte2017,Chng2018,Iso2018,Rao2018} learning and also in the deep learning with advanced structures~\cite{Mills2018,Huembeli2018,Liu2018,Ohtsuki2016,Ohtsuki2017,Morningstar,Liu,Sun,Huembeli,Singh} to classify phases, capture topological features, and locate transition points. An intriguing observation in the supervised learning is that a fixed neural network often works even for systems that were unseen in training. In particular, for the Ising model, the seminal work by Carrasquilla and Melko~\cite{Carrasquilla2017} demonstrated that the phase classifier trained in the square lattices successfully predicted the critical temperature of the unseen triangular-lattice model. Remarkably, the network outputs for different system sizes fell on the same curve in the finite-size-scaling tests.   

We explain these behaviors by solving an analytically tractable model of the neural network that we devise to capture a typical structure emerging in the training of large-scale networks. While the number of hidden neurons is reduced to two in our network, the accuracy of locating the critical temperature is comparable to the previous result with $100$ neurons~\cite{Carrasquilla2017}. It turns out that the information explicitly encoded in the network is the scaling dimension of the order parameter, leading to the interoperability within the class of the same criticality.

\section{Patters in the trained networks}

\begin{figure}
\includegraphics[width=0.48\textwidth]{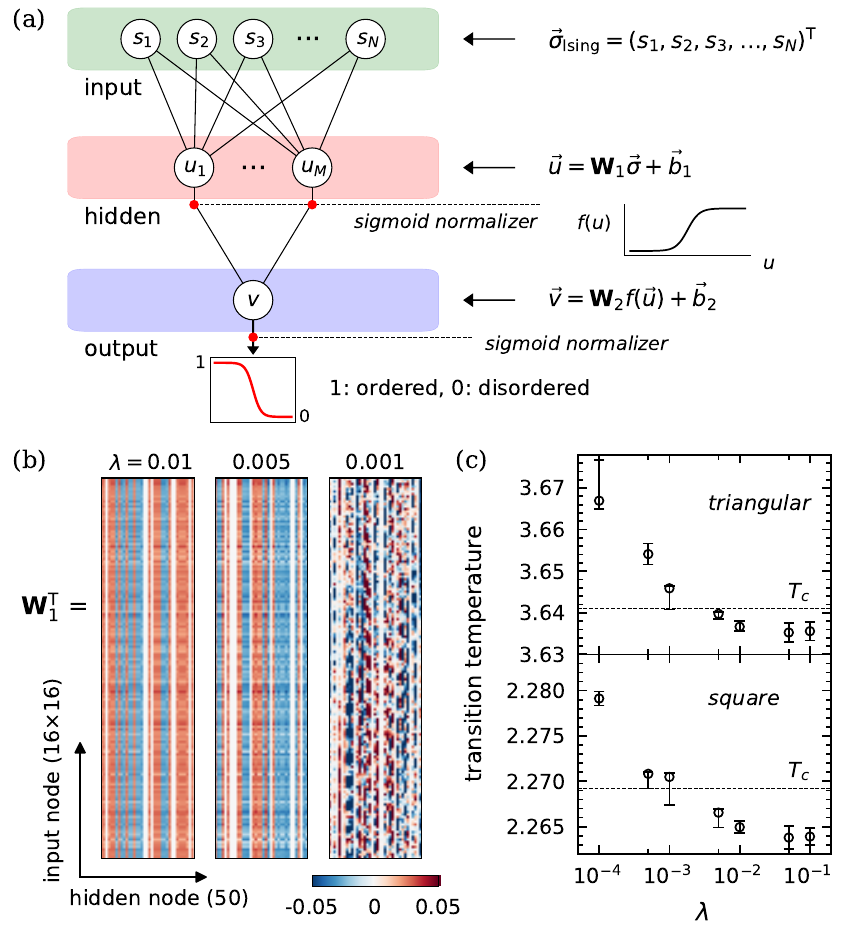}
\caption{Neural network as a phase classifier for the Ising model. (a) The schematic diagram of the signal processing. (b) The examples of the weight matrix $\mathbf{W}_1$ at different values of the regularization strength $\lambda$. (c) The transition temperature as a function of $\lambda$ predicted by the $50$-unit networks which are trained in the square lattices and then applied to the square and triangular lattices for interoperability tests.}
\label{fig1}
\end{figure}

Let us first show the structure that we observe in the network trained in the square lattices (see Fig.~\ref{fig1}). We consider a typical fully connected feed-forward network with a single hidden layer of $50$ neurons between input and output where the sigmoid function normalizes the activation signals. The network is trained by assigning zero (one) to the desired output for the disordered (ordered) phase based on the labeled dataset of spin configurations which are sampled from the Monte Carlo simulations~\cite{Wolff1989} with the Ising Hamiltonian $\mathcal{H} = -\sum_{\langle i,j \rangle} s_i s_j$, where $s_i \in \{1,-1\}$ is the spin state at site $i$, and $\langle i,j \rangle$ runs over the nearest neighbors. The training dataset is prepared at various temperatures around the exact critical temperature $T_c=2/\ln(1+\sqrt{2})$. We use \mbox{\textsc{tensorflow}}~\cite{tensorflow} to minimize the cross entropy with the $L_2$ regularization to avoid overfitting. Details are given in Appendix~\ref{sec:L2reg}.

Two features are notable from the link weights between the input and hidden layers as exemplified in Fig.~\ref{fig1}(b). First, a hidden neuron tends to receive a signal accumulated with almost constant weights of either sign, suggesting that the input $\{s_i\}$ is reduced to its sum $\propto \pm \sum_i s_i$. This is consistent with the activation patterns of the hidden neurons observed previously~\cite{Carrasquilla2017,Suchsland2018} and the concept of the toy model~\cite{Carrasquilla2017}. Second, there are neurons found effectively unlinked with vanishing weights, implying that the size of the hidden layer can be even smaller. 

These features are robust at the regularization strength $\lambda > 0.001$ for all system sizes examined. The structure partly survives at $\lambda = 0.001$, while it fades away as $\lambda$ gets weaker (see Fig.~\ref{fig1s} in Appendix~\ref{sec:L2reg}). We find that the prediction of the transition temperature is consistent at $\lambda > 0.001$ for the networks that are trained in the square lattices and examined also in the triangular lattices (see Fig.~\ref{fig1}(c)). In contrast, as $\lambda$ gets weaker, the accuracy becomes inconsistent in this interoperability test between the different lattices.

\section{Learning the finite-size scaling with the two-unit network model}

Inspired from these observations, we propose a minimal network model by having only two neurons in the hidden layer. The one is linked from the input with a positive constant weight, reading $y \propto \sum_i s_i$, while the other is associated with the opposite sign, reading $-y$. We need a pair of them to learn the $\mathbb{Z}(2)$ symmetry of the Ising model, which is in contrast to the previous toy model~\cite{Carrasquilla2017}. Thus, we write the weight matrix $\mathbf{W}_1$ for the links and the bias vector $\vec{b}_1$ for the hidden neurons as 
\begin{equation}
\mathbf{W}_1 = \frac{1}{N}
\begin{pmatrix}
 1 &  1 & \cdots &  1 \\
-1 & -1 & \cdots & -1
\end{pmatrix}
,\quad
\vec{b}_1 = -\mu 
\begin{pmatrix}
1 \\ 1
\end{pmatrix},
\end{equation}
where $N$ is the number of lattice sites, and $\mu$ is a bias parameter to be determined by training. The outgoing signals from the hidden layer are normalized through the sigmoid function $f(x) = \frac{1}{2}(1+\tanh\frac{x}{2})$. 

We treat the two signals delivered from the hidden to output layer on an equal footing by setting the second weight matrix as $\mathbf{W}_2 = 4\Lambda(1,1)$ with the bias $b_2 = -2\Lambda$ on the output neuron. This choice of $\mathbf{W}_2$ completes the $\mathbb{Z}(2)$ symmetry of the network output for the Ising spin inputs. Being activated also with the sigmoid function, the final output is then written as $q_k =[1+\tanh(\Lambda z_k)]/2$ for an input $\{s_i^{(k)}\}$ where $z_k = 2[f(y_k-\mu) + f(-y_k-\mu)]-1$ for $y_k \equiv \frac{1}{N}\sum_i s_i^{(k)}$.
The pseudotransition temperature $T^*$ can be typically given by $\langle q \rangle_{T^*}=1/2$ where $\langle\cdot\rangle_T$ denotes an average over the dataset at temperature $T$. 

The coverage of the final output function clarifies our motivation for having the common prefactor $\Lambda$ in $\mathbf{W}_2$ and $b_2$. For a value of $\mu$ that is not small, the signal after passing through the network $\mathbf{W_2}$ resides in the range of $(0,4\Lambda)$. Thus, the shift with the bias $b_2 = -2\Lambda$ is a natural choice to make the final output of the sigmoid function cover the full range of $(0,1)$ [see Fig.~\ref{fig1}(a)] as required for successful learning of the phase transition. 

The training of our two-unit network is done by minimizing the cross entropy, 
\begin{equation}
\mathcal{L} = -\int^{T_u}_{T_l} dT\int^1_0 dy \, \rho_T(y) \left[ p_1 \ln q + p_0 \ln (1-q) \right],
\label{eq:entropy}
\end{equation}
where the reference classifier is given by the Heaviside step function $\Theta(x)$ as $p_0 \equiv \Theta(T-T_c)$ and $p_1=1-p_0$, the network output is denoted by $q\equiv q[z(y,\mu),\Lambda]$, and $\rho_T(y)$ is the density of training data giving $y$ at $T$.

Treating this learning problem analytically, we find that an interesting system-size dependence is encoded in the network parameters $\Lambda$ and $\mu$. For simplicity, we approximate $\rho_T(y) \propto \delta (y-m)$ with the order parameter $m\equiv \langle |y| \rangle_T$ by ignoring the fluctuations in the input dataset of $y$. The minimization of $\mathcal{L}(\Lambda,\mu)$ then leads to the following coupled equations,
\begin{eqnarray}
\int^{T_u}_{T_l} dT \left[ 2 p_{1/2} z_m  - z_m \tanh (\Lambda z_m) \right] &=& 0, \label{eq:dLdLam} \\
\int^{T_u}_{T_l} dT \frac{\partial}{\partial\mu}\left[ 2 p_{1/2} z_m - \frac{1}{\Lambda} \ln \cosh(\Lambda z_m)  \right] &=& 0, \label{eq:dLdmu}
\end{eqnarray}
where $z_m\equiv z(m,\mu)$, and $p_{1/2} = 1/2-p_0$. The precise bounds of $(T_l,T_u)$ are irrelevant if it is wide enough because the integrands vanish for a large $|\Lambda z_m|$. Thus, the criterion $|z_m| \lesssim 1/\Lambda$ allows us to define the effective bounds as $T^*\pm\delta T$ centered at the pseudotransition temperature $T^*$ where $z_m(T^*)=0$. Below we show that the effective range of a significant $T$ is comparable to the critical window that scales as $L^{-1/\nu}$ with the length scale $L$ of the system, which becomes essential to understand the behavior of the trained parameters of $\Lambda_L$ and $\mu_L$.

In the area of a small $z_m$ around $T^*$, we may express $z_m$ for $m$ as $z_m \simeq \frac{3}{8}m_*(m-m_*)$ where $m_* \equiv m(T^*)$. In the transition area, we can also replace $m$ by its finite-size-scaling ansatz $m_L = L^{-\Delta_\sigma} \tilde{m}[(T-T_c)L^{1/\nu}]$ with the scaling dimension $\Delta_\sigma \equiv \beta/\nu$, where $\tilde{m}(x)$ is a scale-invariant function. Then, by expanding Eq.~(\ref{eq:dLdLam}) for $z_m$, we can simply write down the leading order behavior of $\Lambda_L$ as
\begin{equation}
\Lambda_L \sim L^{2\Delta_\sigma} \cdot \frac{\int_{x_-}^{x_+} p_{1/2} \tilde{m}_*[\tilde{m}(x)-\tilde{m}_*]dx}{\int_{x_-}^{x_+} \tilde{m}_*^2[\tilde{m}(x)-\tilde{m}_*]^2dx},
\end{equation}
where $x_\pm = (T^*_L-T_c\pm\delta T_L) L^{1/\nu}$. This reduces to $\Lambda_L \sim L^{2\Delta_\sigma}$ when $x_\pm \sim O(1)$, which is indeed confirmed in Eq.~(\ref{eq:dLdmu}). Through the similar procedures, one can also write down the scaling solution of Eq.~(\ref{eq:dLdmu}) as 
\begin{equation}
T^*_L - T_c \sim - L^{-1/\nu}\tilde{m}_* \int^{x_+}_{x_-} \tilde{m}(x)dx + 2 \tilde{m}_*^2 \delta T_L ,
\end{equation}
where $\Lambda_L$ is replaced by $L^{2\Delta_\sigma}$. This holds when $T^*_L-T_c \sim L^{-1/\nu}$ and $\delta T_L \sim L^{-1/\nu}$, or equivalently $x_\pm \sim O(1)$. For $\mu_L$, the equation $z[L^{-\Delta_\sigma}\tilde{m}(T^*_L),\mu_L]=0$ leads to the asymptotic behavior of $\mu_L - \ln 3 \sim L^{-2\Delta_\sigma}$.

\begin{figure}
\includegraphics[width=0.48\textwidth]{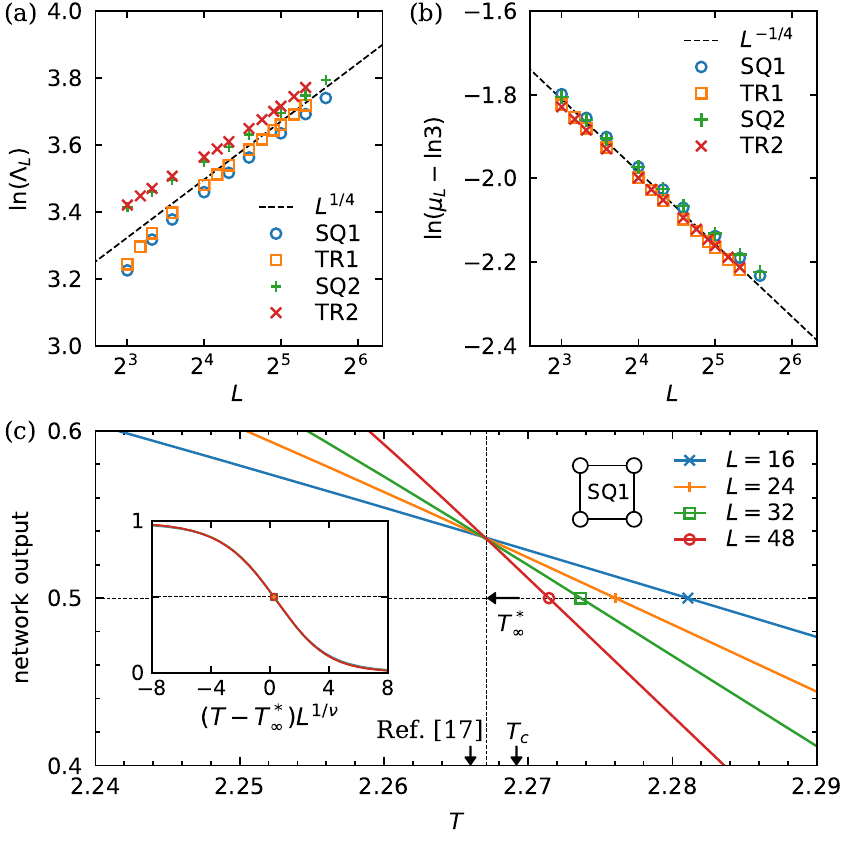}
\caption{Learning with the two-unit neural network. The system-size dependence of the network parameters (a) $\Lambda_L$ and (b) $\mu_L$ trained in the square (SQ1,SQ2) and triangular (TR1,TR2) lattices for $T/T_c \in [0.5,1.5]$ (SQ1,TR1) and $[0,2]$ (SQ2,TR2). (c) The comparison of the transition point predicted by SQ1, the estimate with the $100$-unit network~\cite{Carrasquilla2017}, and the exact $T_c$. The inset shows the scaling collapse of the network outputs with the exponent $\nu=0.94$.}
\label{fig2}
\end{figure}

\section{Numerical verifications}

We numerically verify the behavior of $\Lambda_L \sim L^{2\Delta_\sigma}$ and $\mu_L - \ln 3 \sim L^{-2\Delta_\sigma}$ by performing the learning based on the Monte-Carlo datasets. Specifically, we construct the input data distribution $\rho_T(y)$ by employing the Wang-Landau sampling method for energy and magnetization~\cite{Wang2001a,Wang2001b,Landau2004} (see Appendix~\ref{sec:WL}). This allows us to compute Eq.~(\ref{eq:entropy}) directly with the predetermined $\rho_T(y)$, which makes the minimization numerically straightforward. Figure~\ref{fig2} presents $\Lambda_L$ and $\mu_L$ obtained in two dimensions (2D) for the different choices of the underlying geometry and temperature range for the learning. In all cases, the trained parameters become increasingly parallel to the lines of $\Lambda_L \sim L^{1/4}$ and $\mu_L - \ln 3 \sim L^{-1/4}$ for the exact exponent of $\Delta_\sigma = 1/8$ as $L$ increases.

\begin{figure}
\includegraphics[width=0.48\textwidth]{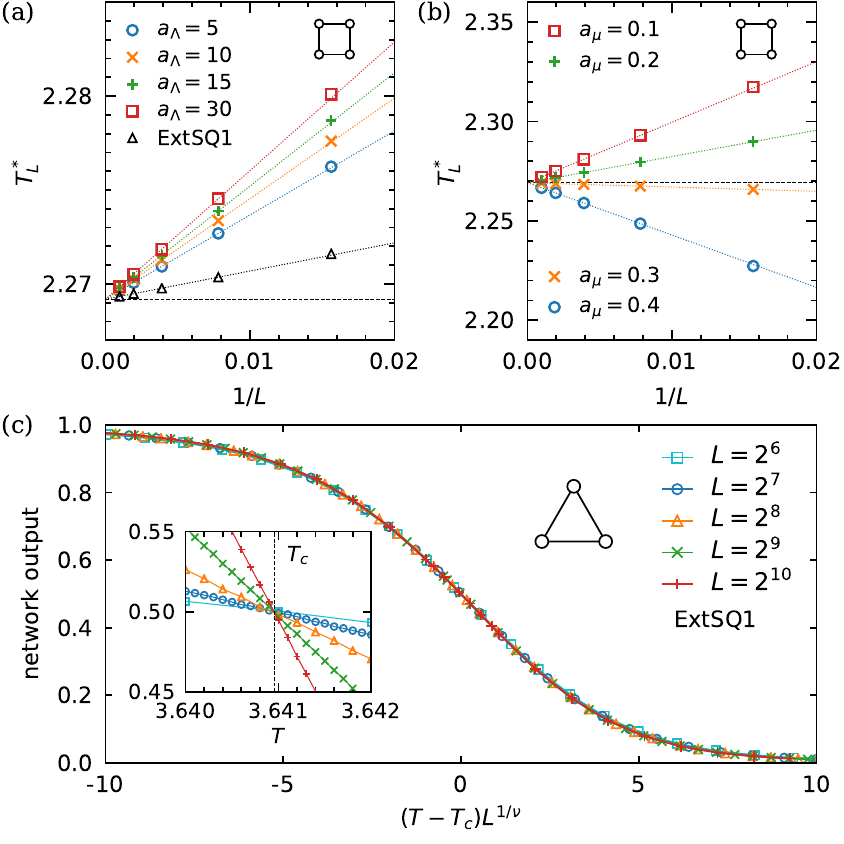}
\caption{Interoperability of the two-unit neural network. At the fixed scaling of $\Lambda_L = a_\Lambda L^{1/4}$ and $\mu_L = \ln 3 + a_\mu L^{-1/4}$, the consistency of finding $T_c$ (dashed line) is examined by varying (a) $a_\Lambda$ at $a_\mu=0.25$ and (b) $a_\mu$ at $a_\Lambda=15$ for the inputs prepared in the square lattices. The network ExtSQ1 is associated with $(a_\Lambda,a_\mu)$ extrapolated from SQ1 trained in the square lattices. (c) The finite-size-scaling test of the outputs of ExtSQ1 for the inputs from the triangular lattices. The exact values of $T_c = 4/\ln 3$ and $\nu=1$ are used. The error bars (not shown) are much smaller than the symbol size. }
\label{fig3}
\end{figure}

It turns out that although we have only two neurons in the hidden layer, the transition point located in our two-unit network is as accurate as the previous estimate with $100$ hidden neurons~\cite{Carrasquilla2017}. In Fig.~\ref{fig2}(c) showing the outputs of the network SQ1 trained and examined in the square lattices, the extrapolation from $T^*_L$ finds $T^*_\infty = 2.267(1)$ with the exponent $\nu = 0.94(2)$ which agrees well with the previous estimate of $T_c = 2.266(2)$ with $\nu = 1.0(2)$~\cite{Carrasquilla2017}. Also, the location of $T^*_\infty$ is at the crossings between the curves of different $L$'s, leading to the scale invariance in the network outputs at the transition temperature. 

The deviation from the exact $T_c$ is possibly due to the finite-size effects of the systems accessible in the learning which are apparent in $\Lambda_L$ and $\mu_L$ at small $L$'s. Since we now know from the analytic results that $\Lambda_L$ and $\mu_L$ should scale asymptotically with the exponent $2\Delta_\sigma$, we may try to remove the finite-size effects by extrapolating the network parameters as $\Lambda_L = a_\Lambda L^{1/4}$ and $\mu_L = \ln 3 + a_\mu L^{-1/4}$ with the expected exponent $\Delta_\sigma=1/8$. We have observed that this parametrization provides $T^*_\infty \approx 2.269$ which is very close to the exact value of $T_c$ (see Fig.~\ref{fig2s} in Appendix~\ref{sec:Tc}).

An important implication of our analytic results is that the essential information encoded by the learning is only the exponent $\Delta_\sigma$ of the critical behavior. Thus, after the training is done, one may not be able to distinguish the networks by the system-specific properties of the training datasets, such as an underlying lattice geometry and a location of $T_c$, as long as they are in the same universality class. This implies that one can actually use the network trained in the square lattices for the prediction with the data in the triangular lattices, explaining the previous observation with the $100$-unit network in Ref.~\cite{Carrasquilla2017}.

Figure~\ref{fig3} shows that locating the precise $T_c$ in the thermodynamic limit is not affected by the training-specific values of $(a_\Lambda,a_\mu)$ when $\Delta_\sigma$ is fixed. For these tests, the input datasets are prepared for the systems with the sizes up to $L=1024$ in the Monte Carlo simulations (see Appendix~\ref{sec:Wolff} for details). The interoperability between the square and triangular lattices is more directly examined in Fig.~\ref{fig3}(c) by using the network ExtSQ1 with $(a_\Lambda,a_\mu)$ being extrapolated from the SQ1 parameters trained in the square lattices. Testing with the data of the triangular lattices shows an excellent scaling collapse at the exact values of $T_c = 4/\ln 3$ and $\nu=1$. The explicit use of SQ1 for small $L$'s provides $T_c \approx 3.637$ (see Fig.~\ref{fig2s} in Appendix~\ref{sec:Tc}), which is also comparable to the $100$-unit network estimate of $T_c = 3.65(1)$~\cite{Carrasquilla2017}.  

\begin{figure}
\includegraphics[width=0.48\textwidth]{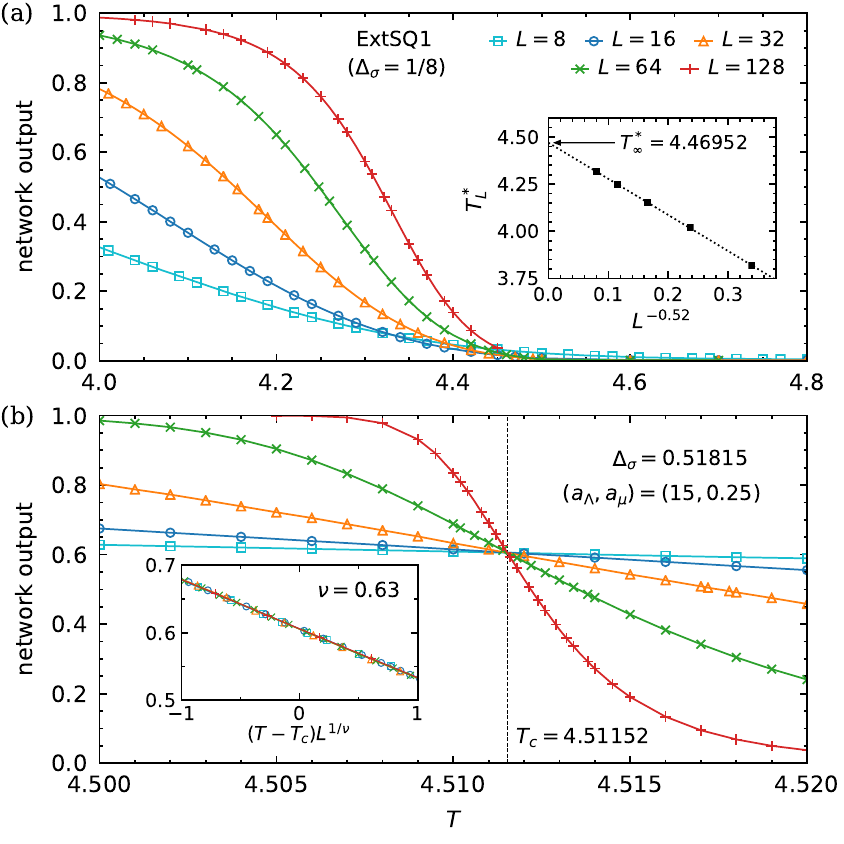}
\caption{Test of the 3D Ising model with the extrapolated two-unit networks. ExtSQ1 is used in (a), examining the behavior of the pseudotransition points $T^*_L$ and the crossing point (not existing) between the output curves. (b) The finite-size-scaling test of the outputs of the network made by using the previous 3D Ising estimate of $\Delta_\sigma=0.51815$~\cite{ElShowk2014}.}
\label{fig4}
\end{figure}

Finally, we discuss what happens in practice when the neural network operates on the system with an exponent mismatch. Figure~\ref{fig4}(a) shows the case where the network ExtSQ1 with the 2D exponent is applied to the inputs given in three-dimensional (3D) cubic lattices. Interestingly, the pseudotransition temperatures $T^*_L$ show a clean power-law convergence to reach $T^*_\infty \approx 4.4695$. It is not a precise $T_c$ and with a wrong exponent, but one might say that it is still not too far from the known $T_c$. However, it clearly loses a scale-invariant point of the output curves, and thus a finite-size scaling test is failed. On the other hand, a network parametrized with the known 3D exponent $\Delta_\sigma = 0.51815$~\cite{ElShowk2014} provides $T_c = 4.51152(1)$ with $\nu=0.63$ (see Fig.~\ref{fig4}(b)) which is in excellent agreement with the previous Monte Carlo estimates~\cite{Hasenbusch2010}.

\section{Conclusions}

In conclusion, we have shown that the minimal binary structure with two neurons in the hidden layer is essential in understanding the accuracy and interoperability of a neural network observed in the supervised learning of the phase transition in the Ising model. We have found that the scaling dimension of the order parameter is encoded into the system-size dependence of the network parameters in the learning process. This allows the conventional finite-size-scaling analysis with the network outputs to locate the critical temperature and, more importantly, demonstrates how one trained neural network can work for different lattices of the same Ising universality.

Explainable machine learning aims to provide a transparent platform that allows an interpretable prediction which is crucial for the applications that require extreme reliability. In the learning of classifying the phases in the Ising model, we have attempted downsizing the neural network to reveal a traceable structure which turns out to be irreducibly simple and yet not to lose its performance. This suggests a necessity of further studies to explore interpretable building blocks of machine learning in a broader range of physical systems.     

\begin{acknowledgments}
This work was supported from the Basic Science Research Program through the National Research Foundation of Korea funded by the Ministry of Education (NRF-2017R1D1A1B03034669) and also from GIST Research Institute (GRI) grant funded by the GIST in 2018.
\end{acknowledgments}

\appendix

\section{Numerical training of the $50$-unit neural network with the $L_2$-regularization}
\label{sec:L2reg}

For the numerical training of the $50$-unit neural network, we construct the loss function $\mathcal{L}(\mathbf{W}_1,\mathbf{W}_2,\vec{b}_1,\vec{b}_2;\lambda)$ by combining the cross entropy and regularization terms (for the overview, see Ref.~\cite{Nielson2015}) as
\begin{eqnarray*}
\mathcal{L} = &-&\frac{1}{n_\mathrm{data}} \sum_{i=1}^{n_\mathrm{data}} \left[ p \ln q + (1-p) \ln (1-q) \right] \\
&+& \frac{\lambda}{4} \sum_{l=1,2} \Vert\mathbf{W}_l\Vert_\mathrm{F}^2, 
\end{eqnarray*}
where the reference classifier $p$ is set to be $1$ if the temperature of the input $\vec{\sigma}_T$ is below $T_c$ and $0$ otherwise, the network output $q$ is a function of the parameter set $(\mathbf{W}_1,\mathbf{W}_2,\vec{b}_1,\vec{b}_2)$ and the input $\vec{\sigma}_T = \{s_1,\ldots,s_N\}$, and the last term is the $L_2$-regularization with the strength $\lambda$ which helps to avoid overfitting. The training dataset includes $1800$ spin configurations per temperature sampled with the spin-up-down symmetry being imposed in the Monte Carlo sampling processes. The spin configurations are sampled for $229$ temperatures regularly spaced with step size $0.01$ in the range of $(0.5T_c,1.5T_c)$, giving the total number of the training data $n_\mathrm{data} = 412200$. The minimization is performed by using the Adam optimizer implemented in \textsc{tensorflow}~\cite{tensorflow}, and the learning proceeds with the entire training dataset during $30000$ epochs at the learning rate $0.0005$. The training is done in the $L \times L$ square lattices for $L=16,20,24,32,40$.

\begin{figure*}
\includegraphics[width=0.95\textwidth]{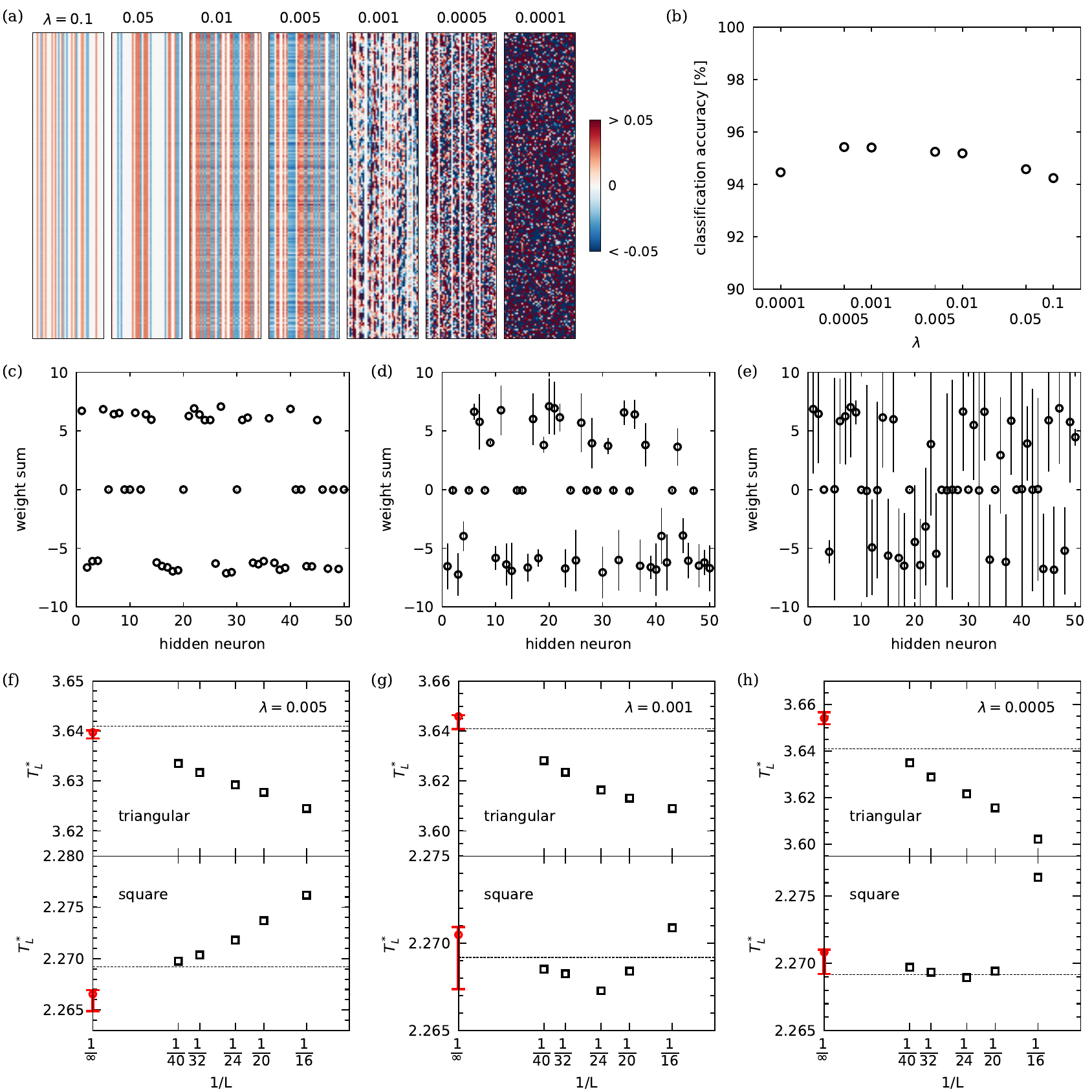}
\caption{Dependence of the training of a neural network on the $L_2$-regularization strength $\lambda$. (a) The visualization of the weight matrix $\mathbf{W}_1^\mathrm{T}$ of the 50-unit network trained in the $16 \times 16$ square lattices. (b) The success percentage in the phase classification test given as a function of $\lambda$. The sum of the weights of the incoming links to a hidden neuron is plotted for (c) $\lambda=0.005$, (d) $0.001$, and (e) $0.0005$, where the length of the bar indicates the magnitude of the partial sum of the weights having the sign opposite to the total weight sum. The panels (f)-(h) present the predictions of the transition temperature in the square and triangular lattices at $\lambda=0.005,0.001,0.0005$.}
\label{fig1s}
\end{figure*}

Figure~\ref{fig1s}(a) visualizes the weight matrix $\mathbf{W}_1$ of the neural network trained at various values of $\lambda$ ranging from $0.1$ to $0.0001$. In the typical validation test of the phase classification with the reserved test dataset of $200$ configurations per temperature in the same range, we have very similar success percentages of about $95\%$ for all $\lambda$'s examined, while slightly higher percentages are found at $0.0005 \le \lambda \le 0.01$ [see Fig.~\ref{fig1s}(b)]. However, this simple classification test does not fully validate the actual accuracy and performance of the network for our purpose of investigating its ability to predict the transition temperature and interoperability with different underlying lattice geometries. 

The direct tests of finding the transition temperature with the networks trained in the square lattices are performed with the datasets separately prepared in the triangular and square lattices. The performance shown in these tests seems to be closely related to the existence or non-existence of the plus-minus structure of the weight observed in the weight matrix $\mathbf{W}_1$ which turns out to undergo a crossover from a structured to unstructured one around $\lambda=0.001$. This crossover does not change with the size of the system that we have examined. One way to notice the change of the visibility of the structure is to look at the weight sum of the incoming links of a hidden neuron as exemplified in Figs.~\ref{fig1s}(c)-\ref{fig1s}(e). At $\lambda=0.005$,  the plus-minus structure is very clear since all contributing weights to a hidden neuron are of the same sign. While the weight sums are still well separated into plus, minus, and zeros, defects start to appear at $\lambda=0.001$, which is shown by the finite length of the bar in Fig.~\ref{fig1s}(d) that indicates the contribution of the weights with the opposite sign to the total sum at a given neuron. At the weaker regularizations of $\lambda \le 0.0005$, the length of the bar tends to get larger to be comparable to the magnitude of the weight sum. The behavior of the pseudotransition temperatures differs as well at these $\lambda$'s as shown in Fig.~\ref{fig1s}(f)-\ref{fig1s}(h). The mark at $L=\infty$ is from the extrapolation with the last three points while the error bars indicate the combined uncertainty of the three- and four-point fittings. Up to $\lambda = 0.005$ of having the clear structure in $\mathbf{W}_1$, the system-size extrapolation with $1/L$ is very consistent. At $\lambda=0.001$, the finite-size behavior becomes severe, and below $\lambda = 0.0005$, the accuracy of predicting the transition temperature in the triangular lattices becomes poor as $\lambda$ further decreases, implying that the overfitting to the reference of a step-function-like classification may have occurred during the training with the data in the square lattices at such small $\lambda$.

\section{Preparation of the input dataset of spin configurations and magnetizations}
\label{sec:Wolff}

The Monte Carlo simulations with the Wolff cluster update algorithm~\cite{Wolff1989} is used to produce the input dataset for training and testing. The spin configurations and magnetizations are sampled at every $N/\langle N_c \rangle$ cluster flip, where $N$ and $\langle N_c \rangle$ are the number of the lattice sites and the average cluster size, respectively. In the measurement of the output of the extrapolated two-unit network, the first $10000$ samples are thrown away during the thermalization, and then $30$ bins of $100000$ samples are used for the measurements. The error bars are estimated by using the jackknife method, but it turns out that they are much smaller than the symbol sizes in all plots and thus are not shown in the figures of the main text.

\section{Preparation of $\rho_T(y)$ to train the two-unit network model}
\label{sec:WL}

We employ the two-parameter Wang-Landau sampling method~\cite{Wang2001a,Wang2001b,Landau2004} to generate the joint density of states $g(E,M)$ of the Ising model by following the standard procedures (for instance, see Ref.~\cite{Kwak2015}, and the references therein). The variables $E = \sum_{\langle i,j\rangle} s_i s_j$ and $M=\sum_i s_i$ cover all possible values of the energy and total magnetization. In all sizes of the system examined, the flatness criterion of the histogram is set to be larger or equal to $0.9$, and the stopping criterion of the modification factor is given as $\ln f < 10^{-8}$. The two-parameter Wang-Landau calculations are known to consume a huge amount of computational time, but still we have obtained $g(E,M)$ up to $L=48$ ($L=40$) in the square (triangular) lattices, where the largest calculation took about four months on a single 3.4 GHz Xeon E3 processor. Note that we have obtained a single set of $g(E,M)$ for each system within our computational resources, and thus the curves in Fig.~\ref{fig2} have been given without error bars. Once the joint density of states $g(E,M)$ is obtained, the distribution function $\rho_T(y)$ of the magnetization $y$ to be used to evaluate the two-unit network can be computed at any temperature $T$ as 
\begin{equation}
\rho_T(y) |_{y=M/N} = \frac{\sum_E g(E,M) \exp[JE/k_\mathrm{B}T]}
{\sum_{E,M} g(E,M) \exp[JE/k_\mathrm{B}T]}, 
\end{equation}
where the ferromagnetic coupling $J$ and the Boltzmann constant $k_\mathrm{B}$ are set to be unity. 

\begin{figure}
\includegraphics[width=0.48\textwidth]{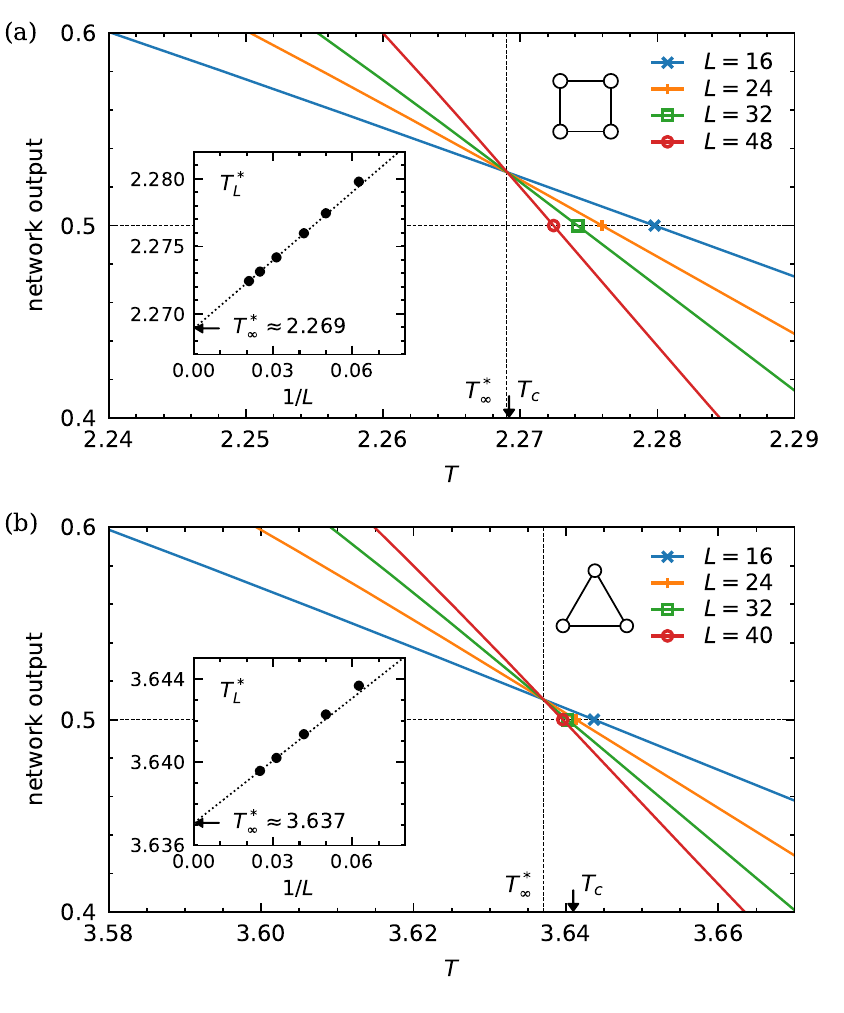}
\caption{Locating the transition temperature with the two-unit networks, ExtSQ1 (a) in the square lattices and SQ1 (b) in the triangular lattices. The mark $T^*_\infty$ indicates the pseudotransition point extrapolated in the thermodynamic limit which is identified in the insets as $T^*_\infty \approx 2.269$ (a) in the square lattices and $T^* \approx 3.637$ (b) in the triangular lattices. The arrows with $T_c$ indicate the location of the exact critical temperature.}	
\label{fig2s}
\end{figure}

\section{Supplemental figures of locating the transition point}
\label{sec:Tc}
	
Figure~\ref{fig2s} displays the supplemental figures of finding the transition temperatures with the two-unit neural networks in the square and triangular lattices. The extrapolated network ExtSQ1 is associated with the parameter set $(a_\Lambda,a_\mu)$ fitted to those of the network SQ1 that was explicitly trained in the square lattices. In the validation the network ExtSQ1 for the transition temperature with the data in the square lattices, the extrapolation of the pseudotransition temperatures ($T^*_L$ along the $0.5$-line of the output) provides $T^*_\infty \approx 2.269$ which agrees very well with the exact value of the critical point $T_c = 2/\ln(1+\sqrt{2})$. On the other hand, in the additional interoperability test of SQ1 with the data in the square lattices, we obtain $T^*_\infty \approx 3.637$ which is also very comparable to the value of $T_c = 4/\ln 3$ of the exact solution in the triangular lattices.

\end{document}